\documentclass[prl,aps,twocolumn,nofootinbib,floatfix,showpacs,superscriptaddress,unsortedaddress]{revtex4}
\usepackage{graphicx}
\usepackage{amssymb}

\begin{document}
\title{Self-sustained oscillations in a Large Magneto-Optical Trap}
\author{G. Labeyrie, F. Michaud and R. Kaiser}
\affiliation{Institut Non Lin\'{e}aire de Nice, UMR 6618, 1361 route des Lucioles,\\
F-06560\ Valbonne.\\}
\date{\today}

\begin{abstract}

We have observed self-sustained radial oscillations in a large magneto-optical trap (MOT), containing up to $10^{10}$ 
Rb$^{85}$ atoms. This instability is due to the competition between the confining force of the MOT and the repulsive 
interaction associated with multiple scattering of light inside the cold atomic cloud. A simple analytical 
model allows us to formulate a criterion for the instability threshold, 
in fair agreement with our observations. This criterion shows that large numbers of trapped 
atoms $N>10^9$ are required to observe this unstable behavior.    
\end{abstract}


\maketitle

A large fraction of the stars in the upper Hertzsprung-Russell diagram present pulsations 
based on an interplay between modulated radiation pressure effects, which tends to 
increase the size of the star, and a collapse based on gravitational forces \cite{Cox}. 
Instabilities also occur in other similar systems such as confined plasmas where a 
long range Coulomb interaction has to be countered by a confining force to avoid an 
explosion of the plasma \cite{Evrard}. These systems are of fundamental importance for 
astrophysics and for controlled fusion and have thus been extensively studied 
in the past. However it is either impossible (in the case of stars) or extremely 
difficult (in the case of confined plasmas) to perform experiments to study the full 
dynamics of such systems where collective effects play a dominant role. On the other 
side, allowing for adequate rescaling, alternative systems can present similar dynamics. 
A variety of interesting collective effects have thus been identified in charged 
colloidal systems \cite{Hansen}. Recently ultra-cold plasmas created by ionizing a cloud of 
laser cooled atoms became subject to increased attention \cite{Simien}. Beyond the possibility 
of studying analogous effects as in astro- and plasma physics, systems with long range 
interactions are known to lead to non-extensive behavior and appropriate scaling laws 
are needed to predict macroscopic properties. Here we show that a large cloud of 
laser cooled atoms is an adequate system to study such collective effects. The 
radiation pressure of the multiply scattered photons in such clouds can indeed 
be related to a long range Coulomb type interaction \cite{Walker}. We thus suggest an 
analogy between the dynamics of a large cloud of cold atoms, astrophysical 
systems and plasma physics.

The effect of multiple scattering on the dynamics of the atoms is well known 
in the community of laser cooling of atoms, as multiple scattering has been a 
major limitation to obtain large phase space densities in cold atomic traps. 
Bose-Einstein condensation (BEC) in dilute atomic vapors has only been achieved after 
switching off all laser fields and using evaporation techniques \cite{BEC}.
More recently, multiple scattering of light in cold atoms has been used to study coherent 
light transport in random media \cite{CBS}.
This has led to 
an investigation of yet unexplored regimes, namely the 
limit of very large number of cold atoms in the presence of quasi-resonant light. 
Here we do not focus on the properties of the scattered light but on the mechanical 
effects of this light on the atoms. We have observed collective instabilities triggered by the repulsive 
interatomic force arising from multiple scattering, and identified a supercritical Hopf bifurcation separating the standard 
stable MOT operation from a yet undescribed unstable regime.

In order to estimate the relevance of plasma physics 
considerations to study 
multiple scattering of light by cold atoms it is worth 
deriving the equivalent of several plasma 
parameters for our system.
The analogy with 
an $1/r^2$ repulsive Coulomb-type 
force \cite{Walker} is obtained from 
evaluating the power scattered by one atom ($P_{scatt}$) 
and deriving the intensity incident $I_{2}$ on a second 
atom via $I_{2} \propto P_{scatt}/(4\pi r^{2})$. 
The resulting radiation pressure force scales as $1/r^2$
and one can thus define an effective 
charge $\tilde{q}$
which depends on the absorption cross sections and laser intensity 
and is typically 
$\tilde{q}\approx 10^{-4} e$\cite{Walker}.
A total interaction 
energy $\tilde{q}V = \frac{N \tilde{q}^2}{4 \pi \epsilon _{0} R} $  larger than the kinetic energy $k_{B}T$ of the particles leads 
to an increased diameter $L=2R$ of the magneto-optical trap when the 
number $N$ of atoms exceeds $\approx 10^5$. Alternatively 
the Debye length $\lambda _{D}=\sqrt{\epsilon_0k_BT/n\tilde{q}^2}$ above which collective effects become important is 
of the order of  $100 \mu m$, well below the typical size of a large MOT (several mm). 
Also, in our experiments the corresponding plasma frequency $\omega_{D}=\sqrt{n\tilde{q}^2/m\epsilon_0}$ is 
slightly larger ($\approx 200Hz$) than relaxation rate of the atomic positions ($\approx 50Hz$). 
We thus expect our cloud to behave as a weakly damped 
plasma. Another interesting 
quantity is the ratio between the nearest neighbor Coulomb interaction and the 
kinetic energy $\Gamma _{Cb}=\frac{\tilde{q}^2}{4 \pi \epsilon_0 a}/k_BT$ with $a \simeq n^{-1/3}$ \cite{Pollock}. 
We estimate this quantity to be smaller than 
unity in our system, excluding thus any crystallization. An important 
aspect of these light induced collective interactions is that the 
effective charge $\tilde{q}$  depends on experimental control parameters, 
allowing for a engineering of the effective charge 
which can be modified 
by orders of magnitude. Finally it might 
be possible to use the high phase 
space densities of a BEC and thus study strongly coupled plasma in the 
degenerate regime \cite{Foldy} as expected in neutron stars and white dwarfs \cite{Jetzer}.

Our cloud of cold atoms is confined in a MOT using laser-induced 
forces \cite{cooling}. We collect $Rb^{85}$ atoms from a dilute vapor using 
six large $\textit{independent}$ laser beams (beam waist $4cm$, power per beam $P=30mW$) 
thus avoiding the intensity imbalance and feedback mechanism responsible for the instability of ref\cite{Lille}. 
Under standard operating conditions, the 
trapping lasers are detuned from the $F=3\rightarrow F^{\prime }=4$ transition of the $D2$ 
line by $\delta = -3\Gamma $ ($\Gamma /2\pi =6MHz$). 
A magnetic 
field gradient ($\nabla B\approx 10G/cm$) is applied to generate a spatially dependent 
Zeeman shift yielding the restoring force of the trap. A repumping laser on the $F=2\rightarrow F^{\prime }=3$ of 
the $D2$ line is used to control the total number of atoms. We thus 
obtain a MOT with up to $N=10^{10}$ atoms (diameter $L=5mm$, $T=80\mu K$)\cite{OptCom}. 
The size and shape of the cloud is monitored by imaging the MOT's 
fluorescence on a cooled CCD. The optical thickness $b$ of the cloud 
at the trapping laser frequency is measured by a photodiode. To 
obtain a time-resolved information on the local density 
of the MOT, we also image a portion of the cloud on another photodiode.

\begin{figure} 
\begin{center} 
\includegraphics[width=8cm]{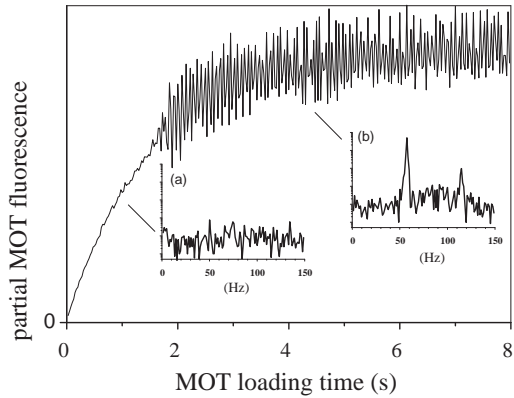} 
\end{center} 
\caption{\label{loading} 
Fluorescence of part of the MOT during a loading 
sequence. Below a critical number of atoms $N_{th}$, the size of the atom cloud
 increases without specific dynamical behavior. Above the threshold $N_{th}$ the cloud 
switches to an unstable mode characterized by periodic 
oscillations in the partial fluorescence signal.
Inserts : Fourier transform of signal, with (a) a flat noise
in the stable regime and (b) distinct oscillations in the unstable regime.} 
\end{figure}

Fig.\ref{loading} illustrates the onset of spontaneous 
self-sustained oscillation for a sufficiently large number of atoms. We switched on the MOT at $t=0$ and 
monitored the time evolution of the fluorescence from a 
portion of the MOT. This partial fluorescence signal is roughly describing the number of atoms in the 
observed region. 
Starting from $N=0$ at $t=0$, the trap fills with a time constant $\tau=1.45s$ determined by the ambient 
Rb pressure. Below a critical 
number of atoms $N_{th}$, the size of the atom cloud increases with number of trapped atoms\cite{Wieman2} but no specific dynamical 
behavior is observed. 
Above the threshold $N_{th}$ the cloud switches to an unstable behavior characterized by 
periodic oscillations in the partial 
fluorescence signal. These radial oscillations of the cloud are self-sustained 
in the sense that no external modulation of any 
control parameter is present. 

In the inserts of Fig.\ref{loading} are shown Fourier transforms 
of the partial fluorescence. Below 
the instability threshold (insert a) 
a flat noise background is obtained.  
In contrast, in an unstable MOT,
obtained for a larger 
number of atoms, 
distinct oscillation frequencies (insert b) appear, 
with higher harmonic components
indicating 
the non harmonic oscillation of the signal.

\begin{figure} 
\begin{center} 
\includegraphics[width=8cm]{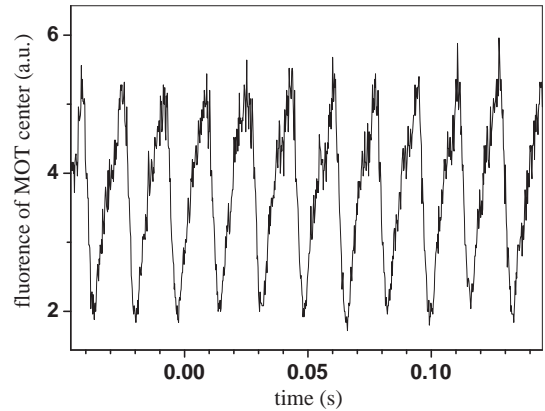} 
\end{center} 
\caption{\label{time} 
Fluorescence of the MOT center. In the unstable regime, periodic oscillations 
appear in the absence of external modulations. } 
\end{figure}

Indeed, the dynamics in the unstable regime can be more complex than a harmonic oscillation, 
as further illustrated in Fig.\ref{time} where 
we detect the fluorescence from the center of the MOT. 
A high contrast modulation of the center fluorescence is observed in this experiment. 
We can speculate that to the fast phase of decrease of the signal corresponds a MOT expansion (decreased density at the center), whereas 
we associate the increasing part of the fluorescence 
to a slower compression phase. 
We observed that the precise shape of this oscillation depends on the 
laser beam alignment and on the monitored region of the MOT. However, the threshold separating the stable from the unstable regime was found to be 
very robust with respect to trap parameters.

\begin{figure} 
\begin{center} 
\includegraphics[width=8cm]{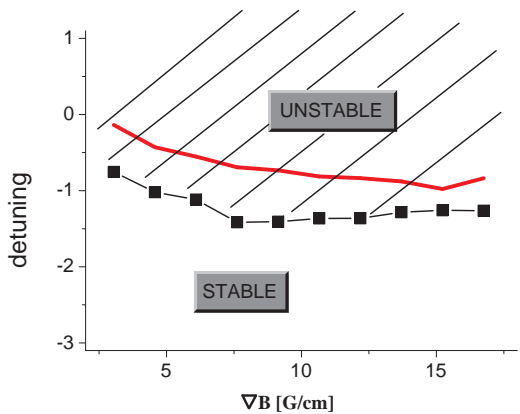} 
\end{center} 
\caption{\label{diagram} 
Phase diagram of self-sustained oscillation : a $(\delta , \nabla B)$ cut in parameter space 
shows the separation between the stable (for larger detuning $\delta$)  and the 
unstable regime: experimental threshold values (squares) and 1-zone model prediction (solid line).} 
\end{figure}

Investigating the MOT at the instability threshold by varying 2 of the control 
parameters of the experiment (detuning, magnetic field gradient), we can map the phase 
diagram shown in Fig.\ref{diagram} (full squares). The solid line corresponds to the theoretical prediction presented at the end 
of this paper. 
As can be seen, the overall behavior is unstable when the trapping laser frequency is 
brought within roughly one natural 
width from resonance. This critical detuning depends here rather weakly on the magnetic field gradient. Note however that the 
measured cloud size and number of atoms {\it at threshold} do vary quite a bit during this experiment (a factor 5 for $N$ and a factor 2 for $L$). 
In addition, we systematically found an optical thickness $b \approx 1$ at the instability threshold. 
However, this is clearly not a sufficient condition for the onset of instabilities, since $b = 1$ is also observed in the stable 
region of Fig.\ref{diagram}.

We also carefully monitored the fluctuations of the total number of atoms when the MOT operates in the unstable regime, 
 as e.g. for the data in Fig.\ref{time}. We found these to be below $2\%$, as in the stable regime. This indicates that, for 
a given set of MOT parameters, the unstable cloud oscillates at fixed $N$. 

To further characterize the transition to instability, we have checked that the amplitude of the oscillation continuously grows from 
zero as the control parameter (detuning or number of atoms in the experiments performed) crosses the threshold value. A 
Fourier analysis of the 
signal across the threshold showed that the instability start at a non zero frequency, which is 
closely related to the natural oscillation frequency of the harmonic trap. Furthermore, no hysteresis was observed despite explicit investigation. All 
these findings are consistent 
with a supercritical Hopf bifurcation.

Already in the stable regime, we observed some clear indications that strongly increasing the number of trapped atoms affects the way the MOT 
operates. As it is well-known, the MOT inflates when atoms are added as a consequence of multiple scattering of light \cite{Walker}. In 
addition to the standard $L \propto N^{1/3}$ law\cite{Walker}, we found for large number of atoms $N>10^9$ a different scaling 
$L \propto N^{1/2}$\cite{Wieman2,taille}. By monitoring the relaxation of the MOT after displacing it from its equilibrium position, we observed 
a cross-over from an over-damped behavior at small $N$ (typical for usual MOTs) to an under-damped behavior at large $N$. We interpret this finding 
as a consequence of the attenuation of the trapping beams inside the cloud, which reduces the friction at the center of the cloud. This could be 
envisioned as a precursor to the instability. Indeed, we found that just below the threshold (i.e. in stable operation), the MOT is systematically 
in the underdamped regime.

To explain the apparition of this new instability, we developed a simple model where the screened compression force of the MOT is competing against 
the repulsive interaction due to multiple scattering of light inside the cloud. We stress that this instability is thus qualitatively different from 
that studied 
in ref\cite{Lille}, where the use of retro-reflected beams introduces the feedback necessary for the instability. The instability process of 
ref\cite{Lille}, which manifests as oscillations of the center-of-mass of the MOT, does not involve the long-range interatomic interactions which drives 
the behavior of our large MOT. 

We propose in the following a very 
simple 1-zone model which exhibits an instability 
threshold. This model amounts to an extremely simplified mean field theory, based however on microscopic expressions for the light forces acting on 
the atoms. A more refined approach, beyond the scope of this paper, could e.g. involve 
hydrodynamical approximations\cite{Pohl}.  

We assume an homogeneous 
density and the size of the
 cloud $L$ is related to the density $n$ via the total number of atoms $N$: $n=N/L^3$. The 
dynamics along one symmetry
 axis ($Ox$) of a probe particle located outside of the cloud (at position $x > R$ from 
the trap center, with a velocity $v$) 
is then governed by the force :
 
\begin{eqnarray}
\label{force}
F\left( x,v \right) & = &    \frac{\hbar k \Gamma }{2} s_{inc} \frac{e^{-b}}{1+ \frac{4(\delta- \mu x -kv)^2}{ \Gamma^2}} \nonumber \\
& & - \frac{\hbar k \Gamma }{2} s_{inc} \frac{1}{1+ \frac{4(\delta+ \mu x +kv)^2}{ \Gamma^2}} \nonumber \\
& & + \eta  \frac{\hbar k \Gamma }{2} s_{inc} \frac{1}{1+ \frac{4\delta^2}{ \Gamma^2}} (1-e^{-b}) (\frac{R}{x})^2 \nonumber \\
\end{eqnarray}

This expression relies on the low intensity Doppler model for the magneto-optical force 
(incident on-resonance 
saturation parameter $s_{inc}$). The first term in this expression is the attenuated 
force of the laser passed through the
 cloud (with the corresponding Zeeman shift $\mu x$ and Doppler shift $kv$), the second 
term corresponds to the non 
attenuated force of the laser propagating in the opposite direction. 
In absence of the $e^{-b}$ attenuation, these two terms give rise to the standard 
cooling (via the opposite Doppler terms $kv$) and
trapping (via the opposite Zeeman terms $\mu x $) of cold atoms.
  The
 last terms is the sum of all binary repulsive interactions which, using Gauss' theorem, 
yields an $1/r^2$ repulsion for a probe 
particle outside the cloud. 
This term can be understood as the radiation pressure originating from the MOT 
with a total radiated power corresponding to attenuation of the 6 laser beams.
Here $\eta$ corresponds to the ratio between the absorption cross section 
of the incident laser frequency and the
 inelastically rescattered photons \cite{Walker}. We now apply this model at the edge 
of the cloud ($x=R$). A linear stability 
analysis, using $x=R+\delta x \ e^{i \omega t}$  around the fixed point $F(R, v=0)=0$ , 
yields the threshold condition for an instability ($Im(\omega)=0$) :
 
\begin{equation}
C\left( \delta , \mu ,b,R \right) = e^{-b} \frac{\delta - \mu R }{1+ \frac{4 (\delta - \mu R)^2}{ \Gamma^2}}
+ \frac{\delta + \mu R }{1+ \frac{4 (\delta + \mu R)^2}{ \Gamma^2}}
=0 \label{seuil}
\end{equation}

We find that for our experimental parameters the threshold is given with a good 
approximation by :
\begin{equation}
\delta + \mu R \approx 0 
\label{criteresimple}
\end{equation}

It should be stressed that in standard MOTs one usually has $\mu R \ll |\delta|$ : for $|\delta|=\Gamma$ and a magnetic field gradient 
$\nabla B=10$ G/cm, condition (\ref{criteresimple}) implies a MOT diameter of 8 mm ! It is only with $N$ in the $10^{10}$ range that such MOT sizes 
can be 
obtained. In this regime, the edge of the cloud is now exploring the nonlinear 
part of the magneto-optical force. It can be shown from the expression of the force that the threshold 
condition (\ref{criteresimple}) corresponds to the passage from a positive to a negative friction 
at the edge of the cloud. Thus, a small velocity fluctuation is amplified instead of damped and the atoms at the edge are kicked away from 
the center 
 of the cloud. The cloud thus expands until its optical thickness drops below a certain level, where the MOT is back to its standard mode of operation 
(weak repulsion) and the atoms are pushed back toward the center. 

In order to further confront our simple 
model to the experiment, we computed the threshold value of the detuning $\delta$ 
using the control parameter $\nabla B$ and 
the $\textit{measured}$ values of the size of the cloud $R$ and its optical thickness $b$ at 
threshold. The result corresponds to the solid line in Fig.\ref{diagram} and gives the correct order of magnitude and behavior for the 
instability threshold. This gives us further confidence about the qualitative validity of our model, as the size of the cloud varies by more than a 
factor 2 along the threshold boundary. Furthermore, both the threshold criterion (\ref{criteresimple}) and the simple picture of the unstable dynamics near 
threshold are confirmed by recent 
results from a more involved model\cite{Pohl2}, which includes the inhomogeneous density distribution inside the cloud (N-zone model).

We have described in this paper the observation of self-sustained oscillation 
in a large cloud of laser cooled 
atoms, arising from the competition between the MOT's confining force and the long range multiple scattering repulsive interaction. 
This new instability process affects the behavior of large MOTs containing more than $10^9$ atoms, a regime increasingly found e.g. in 
experimental setups producing BECs. This 
observation shows that large clouds 
of cold atoms still present a rich dynamics with a variety of yet unexplored 
regimes. A simple 1-zone model has 
been presented which allows to predict the instability threshold and to understand 
the underlying physical mechanism. 
Future possible investigations 
include the forced oscillation regime, the spectroscopy of excitation modes in 
this system, gas-liquid-crystal phase transitions in the degenerate regime and feedback 
mechanisms allowing for stabilization of a large cloud of interacting particles. 
Progress on the theoretical aspects 
of the systems described in this paper include exploiting mean field theory 
and molecular dynamic simulations. 
This should allow for a better understanding of the bifurcation observed 
in our experiment and lead to study 
statistical (thermodynamic) properties across the threshold. If the 
degenerate regime could be reached (or 
inducing similar interactions in a Bose Einstein Condensate) a mean 
field theory based on binary collisions as
 in usual Gross Pitaevskii equation will not be valid due to the 
long range interaction, connecting this system to 
strongly correlated quantum systems.

We thank L. Gil, Ph. Mathias, D. Wilkowksi, S. Balle, F. Bouchet, B. Cessac and 
Y. Elskens for insightful conversations 
and E. Vaujour and M. Renaudat for help at various stages of the experiment. 
We acknowledge financial support by CNRS and PACA region.


\begin{thebibliography}{9}
\bibitem{taille}  The observed $\sqrt N$ size dependence of large MOTs is not yet fully understood. Several physical 
ingredients could possibly yield such a variation : multiple scattering (beyond double scattering), non linear 
spatial dependence of the forces or non uniform atomic density.
\end{thebibliography}
\end{document}